\documentclass[journal]{IEEEtran}

\usepackage{color}
\usepackage{graphicx}
\usepackage{amsfonts}
\usepackage{amsmath}
\usepackage{graphicx}
\usepackage{verbatim}
\usepackage[update,prepend]{epstopdf}
\usepackage{cite}
\usepackage{epsfig}
\usepackage{latexsym}
\usepackage{subfigure}
\usepackage{epstopdf}
\usepackage{verbatim}
\usepackage{units}
\usepackage{amsthm}
\usepackage[longend,ruled]{algorithm2e}
\usepackage{balance}

\usepackage{floatflt}

\usepackage{stfloats}

\usepackage{booktabs}

\newtheorem{definition}{Definition}

\begin{document}
\title{Spectrum Matching in Licensed Spectrum Sharing}
\author{M.~Majid~Butt,~\IEEEmembership{Senior~Member,~IEEE,}~Irene Macaluso,~Eduard~A.~Jorswieck,~\IEEEmembership{Senior~Member,~IEEE,~}\\Julie~Bradford,~Nicola~Marchetti,~\IEEEmembership{Senior~Member,~IEEE,}~and~Linda Doyle

\thanks{Irene Macaluso, Nicola~Marchetti and Linda Doyle are with CONNECT center for future networks, Trinity College Dublin, Ireland. Email:\{macalusoi,~nicola.marchetti,~linda.doyle\}@tcd.ie.}
\thanks{M. Majid Butt is with school of engineering, University of Glasgow, UK. Email: majid.butt@glasgow.ac.uk.}
\thanks{Eduard A. Jorswieck is with Department of Electrical Engineering and Information Technology, TU Dresden, Germany. Email: eduard.jorswieck@tu-dresden.de.}
\thanks{Julie Bradford is with Real Wireless, UK. Email: julie.bradford@realwireless.biz}
\thanks{The project ADEL acknowledges the financial support of the Seventh Framework Programme for Research of the
European Commission under grant number: 619647. We also acknowledge support from the Science Foundation Ireland
under grants No. 13/RC/2077 and No. 10/CE/i853.}

}
\maketitle
\begin{abstract}
Spectrum sharing is one of the promising solutions to meet the spectrum demand in 5G networks that results from the emerging services like machine to machine and vehicle to infrastructure communication. The idea is to allow a set of entities access the spectrum whenever and wherever it is unused by the licensed users. In the proposed framework, different spectrum provider (SP) networks with surplus spectrum available may rank the operators requiring the spectrum, called spectrum users (SUs) hereafter, differently in terms of their preference to lease spectrum, based for example on target business market considerations of the SUs. Similarly, SUs rank SPs depending on a number of criteria, for example based on coverage and availability in a service area. Ideally, both SPs and SUs prefer to provide/get spectrum to/from the operator of their first choice, but this is not necessarily always possible due to conflicting preferences. We apply matching theory algorithms with the aim to resolve the conflicting preferences of the SPs and SUs and quantify the effect of the proposed matching theory approach on establishing preferred (spectrum) provider-user network pairs. We discuss both one-to-one and many-to-one spectrum sharing scenarios and evaluate the performance using Monte Carlo simulations. The results show that comprehensive gains in terms of preferred matching of the provider-user network pairs can be achieved by applying matching theory for spectrum sharing as compared to uncoordinated spectrum allocation of the available spectrum to the SUs.

\end{abstract}

\section{Introduction}
Tremendous increase in data demand and the range of services provided in 5G wireless networks has resulted in various approaches to meet the spectrum demand. As the 5G features in terms of envisioned technologies and service demands dictate, such a trend is not expected to decrease for the foreseeable future. In order to address the corresponding wireless capacity demand, it is required to allocate additional spectrum for 5G communication services. This goal can be reached mainly via three approaches \cite{Morgado:2015}: (i) Clearing (also known as refarming) the spectrum and allocating it to 5G services; (ii) Sharing spectrum between existing incumbents and 5G operators; (iii) Using millimeter wave (30GHz-300GHz) technology.

Refarming is seen as a not-so-straightforward solution, as most of the useful frequency bands are already occupied by other radio services and in general cannot be cleared in a short time frame. Millimetre wave (mm-wave) is concerned with taking advantage of the vast amount of spectrum available in the range of 30 to 300 GHz. Bands at these frequencies have not previously been considered for cellular access, due to rain attenuation, atmospheric absorption, and huge propagation losses compared to lower carrier frequencies. In general, several technological challenges have still to be addressed before millimeter wave technology can be fully integrated in 5G networks; despite this, mm-wave is seen as a promising technology with applications in indoor environments and back-hauling of small cells \cite{Sexton:2017}.

In general, spectrum sharing is seen by national regulators, in both Europe and US, as a viable solution for allocating additional spectrum in a timely fashion, since technologies  that are capable to implement it already exist \cite{Morgado:2015}. In the context of spectrum sharing, cognitive radio has been one of the most popular research approaches \cite{Ashour_TCOM:2016}. The interested reader can refer to \cite{Hossain:2015, Tehrani:2016, Paisana:2014} for survey papers on cognitive radio and spectrum sharing, with a view towards future networks. The author of \cite{Nevokee:2010} provides a survey on the usage of cognitive radio to access TV white spaces. In general, the exploitation of TV white spaces by uncoordinated unlicensed secondary users, implies a lack of quality of service (QoS) guarantees to the secondary users, which has rendered this solution unattractive to mobile network operators \cite{Morgado:2015}.

In \cite{Paisana:2014}, the authors provide a brief but comprehensive description of the main spectrum access techniques, namely geolocation database, beacon signaling, spectrum sensing and cooperative sensing, which have been proposed in the literature for the detection of spectrum holes; the authors then go on to analyse how each of the above mentioned techniques is in turn affected by the radio environmental factors. The authors of \cite{Shokri:2016} investigate the extent to which spectrum sharing in millimeter-wave networks with multiple cellular operators is a viable alternative to traditional dedicated spectrum allocation.


On top of sharing in unlicensed bands, Licensed shared access (LSA) and Citizens Broadband Radio System (CBRS) have been identified as possible solutions for spectrum crunch. LSA and CBRS support sharing of spectrum in under-utilized radar (e.g., marine and less critical aeronautical services) bands between incumbents and one or more secondary network operators for relatively longer periods \cite{Elma:GC2016,Frascolla_VTC:2016,Majid_PIMRC:2016}.

In more detail, LSA is a two tier model for spectrum sharing in licensed bands. The top tier in LSA consists of incumbent users, who own spectrum and have guaranteed protection from the secondary users. It is possible for the incumbents to extract revenue from the under-utilised spectrum they own. The second tier consists of secondary LSA licensees, who can get short term access rights with a guaranteed quality of service to the under-utilised spectrum licensed by incumbents. Protection of the incumbents by sharing in a non-interfering manner is of critical importance. Given that the incumbent activity in these bands is often localized in time and/or space, this leads to the possibility for potential secondary use inside specified areas, or at specific times. Mobile network operators (MNOs) can use (on an exclusive basis) the licensed spectrum owned by other incumbents when and where these incumbents are not using it. In this way, the incumbents are protected from harmful interference and the licensees benefit from the provision of predictable QoS \cite{Mueck:2014, NSN-Qualcomm:2011}. The band under consideration for LSA use is 2.3-2.4 GHz in Europe \cite{Sexton:2017, Morgado:2015}.

CBRS differs from LSA in several respects, the number of tiers being perhaps the most important. In addition to incumbents and high-priority licensed users, a third tier of low-priority users is allowed to access the spectrum  \cite{ETSI:2014, fcc2015,fcc2016}. The third tier in CBRS is known as general authorized access (GAA). GAA users are allocated spectrum resources with no interference protection guarantees, and require active management to ensure that they do not interfere with either tier one or two users. The band foreseen for CBRS deployment is the 3.5GHz band in the USA \cite{Sexton:2017, Morgado:2015}. For the emerging applications in 5G networks and the spectrum sharing opportunities available through LSA and CBRS, novel spectrum allocation techniques need to be developed, due to the specific characteristics of said spectrum sharing regimes.

\subsection{Related Work and Our Contribution}
\label{sect:related_work}
Several mathematical tools from the field of economics, including game theory, auction theory, etc., have been applied to wireless resource allocation problems \cite{Ni_JASC:2012}. Another such tool that has gained momentum recently is matching theory, whose advantages for wireless resource management include the ability to define general preferences that can handle heterogeneous and complex QoS-related considerations; and efficient algorithmic implementations that are inherently self-organizing \cite{Gu:2015, Gu:2016}. Deferred acceptance (DA), introduced in \cite{Gale:1962}, is an  efficient algorithm that can find such a matching. Users and resources make their decisions based on their individual preferences (e.g., QoS metric). The authors of \cite{Gu:2015} provide the first comprehensive tutorial on using matching theory to develop innovative resource management mechanisms in wireless networks, discussing the fundamental concepts of matching theory and a variety of properties that allow definition of several classes of matching scenarios.
Applications of matching theory to wireless networks range from physical layer security systems \cite{Bayat:2013-may} to small cell networks \cite{Pantisano:2013}, from distributed orthogonal frequency-division multiple access (OFDMA) networks \cite{Jorswieck:2011} to heterogeneous cellular networks \cite{Bayat:2014-aug}. \cite{Pantisano:2013} proposes a framework which accounts for interference and performs downlink cell association for a context-aware network in which preferences capture information including application type, hardware size, and physical layer metrics. In \cite{Jorswieck:2011} the authors show that classical schemes such as proportional fair often yield unstable matchings motivating the need to analyze and optimize stable matchings for self-organizing wireless systems. Matching theory is applied in \cite{Zhou_Access:2017} to establish energy-efficient stable matching for device to device pairs and user equipment. For a recent survey of the progress and challenges in application of matching theory to wireless networks, the reader is referred to \cite{Bayat:2016}.

The application of matching theory to spectrum sharing has been widely investigated \cite{Leshem:2012, Leshem:2010, Bayat:2013-aug}. In \cite{Leshem:2012} a one-to-one matching problem is formulated between a set of secondary users and a set of primary users (channels), depending on the rate achievable over such channels. In \cite{Hamza:2017} the authors consider a two-sided one-to-one matching problem, where secondary users relay primary users' data in exchange for spectrum access time; the utility of primary users is the data rate, while for secondary users the authors account for the rate and the power used to help the primary. A similar approach is adopted by \cite{Gao:TMC2016}, where for each pair of matched primary and secondary users, the primary and secondary users' utilities depend on the secondary user's relay power and primary user's spectrum access time reward. \cite{Jorswieck:2013} proposes a novel distributed two-stage resource allocation technique for multiple input multiple output (MIMO) cognitive radio links operating within an environment of multiple multi-antenna primary links. The secondary users are matched to primary users' resources; the preferences of the secondary users depend on both their own channels and the leakage channels, while the preferences of the primary users depend on the interference channels.

All the above mentioned works focus on spectrum sharing in a cognitive radio context, while our work aims at applying matching theory to spectrum sharing frameworks with coordinated access to spectrum, focusing in particular on CBRS. Spectrum resources are provided by a spectrum provider (SP) network and leased to a spectrum user (SU) network. Both SPs and SUs have strict preferences (depending on their individual and possibly heterogeneous requirements) in the order they want to sell and lease spectrum, respectively and these preferences may not always match. In this work, we first develop a general framework for spectrum sharing between the SPs and the SUs and apply the matching theory algorithms to the spectrum allocation problem. Then, we present a specific example within the CBRS framework and show how matching theory helps both SPs and SUs to best match their priorities in terms of spectrum access on the available resources. One of the main advantages of using matching theory for spectrum allocation is the possibility for both SPs and SUs to express preferences that can embed heterogeneous and complex considerations not necessarily or exclusively related to technical requirements. The ability of matching preferred SP-SU pairs is quantified numerically using Monte Carlo simulations. The results show the gain provided by the matching theory algorithms in terms of probability of each SU getting spectrum from the SP of its first choice.

The rest of this paper is organized as follows. Fundamentals of matching theory are introduced in Section \ref{sect:matching theory}. We discuss spectrum sharing scenarios in Section \ref{sect:spectrum sharing} and performance for these scenarios is quantified numerically in Section \ref{sect:performance evaluation}. Section \ref{sect:conclusions} concludes with the summary of main results.

\section{Fundamentals of Matching Theory}
\label{sect:matching theory}
First, we formally define the terms SPs and SUs:
\begin{definition}{Spectrum Provider Network (SP)} is a service provider that provides surplus spectrum to the secondary market on discrete spectrum allocation instants after meeting spectrum demands of its own services.
\end{definition}
\begin{definition}{Spectrum User Network (SU)} is a service provider that provides various services, but does not own any licensed spectrum and requests spectrum from the primary market when needed.
\end{definition}
We consider a network with $N$ SUs and $M$ SPs. Based on various factors involved (we come back to this point later), the SUs and SPs have their own priorities for spectrum allocation from/to SPs and SUs, which may not coincide always. Favorable pairing of SPs and SUs can help both SPs and SUs use spectrum more efficiently.

Based on network spectrum sharing constraints, two sharing scenarios can be defined which are explained below.
\subsection{One-to-One Scenario}
\begin{figure}[!t]
\centering
\includegraphics[width=1.5in]{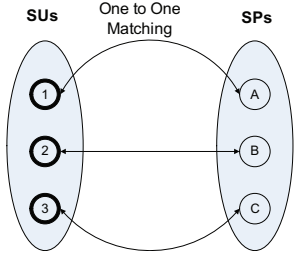}			
\caption {One-to-one scenario for $M=3$ and $N=3$.}
\label{fig:OTO}
\vspace{-0.3cm}
\end{figure}

In the One-to-One scenario, every SP wants to lease spectrum to only one SU. This could be attributed to various reasons: avoiding complex interference management across various SUs with different system requirements or just a case of not allocating spectrum to two SUs competing for business. This implies if $N>M$, some of the SUs end up having no spectrum. We use this scenario for bench marking purposes. This scenario has been illustrated in Fig.~\ref{fig:OTO}.

To maximize SU's probability of getting spectrum from the SP of its first preference, we apply stable marriages matching algorithm \cite{Gale:1962}, which has been applied to solve similar problems in wireless networks \cite{Gao:TMC2016}.
We first define some terms used in this work.
Denoting set of SPs and SUs by $\mathcal{M}=\{1,\dots,M\}$ and $\mathcal{N}=\{1,\dots,N\}$, we define the following terms.
\begin{definition}{\textbf{Two sided matching market:}} A two sided matching market is a market consisting of two disjoint
sets of operators, where an operator on one side can be matched with only one operator on the other side.
\end{definition}
\begin{definition}{\textbf{One-to-One Matching:}} A one-to-one matching between two disjoint sets $\mathcal{M}$ and $\mathcal{N}$ can be represented by a one-to-one correspondence $\mu(.)$, where $m \in \mathcal{M}$ is mapped to $n\in \mathcal{N}$ if and only if $n$ is also mapped to $m$ \cite{Gao:TMC2016}.
\end{definition}
Let us denote one-to-one matching from $m$ to $n$ by $\mu(m)=n$; and $n$ to $m$ by $\mu(n)=m$.
\begin{definition}{\textbf{Preference:}} The preferences of an SP $m$ is defined in terms of an ordered list in the decreasing preference, $P(m)$, on the SU set $\mathcal{N}$ such that
\begin{equation}
P(m)=(n_1\dots,n_Q).
\end{equation}
where $Q\leq N$.
\end{definition}
Similarly, preferences $P(n)$ of SUs can be defined as,
\begin{equation}
P(n)=(m_1,\dots,m_L)~.
\end{equation}
where $L\leq M$.
Note that cardinality of elements in preference sets $P(m)$ and $P(n)$ can be less than $N$ and $M$, respectively as it is not mandatory for every SP and SU to have all the SUs and SPs in their preference list. For all the scenarios considered in this work, we assume that both SUs and SPs have strict preferences and the preferences are known.

We quantify the performance of the proposed algorithms and scenarios using a quantitative measure called matching success $S_{n,i}$ for an SU $n$ given by,
\begin{equation}
\begin{split}
 &S_{n,i} = \\
 &\frac{\mbox{No. of times spectrum from SP with preference $i$ allocated}}{\mbox{Number of times spectrum requested}}
\end{split}
\end{equation}
Our objective is to maximize $S_{n,1}$ and $S_{n,2}$ for the SU $n$ given that the preferences of the SP are fixed\footnote{The goal is to maximize $S_{n,1}$. However, if the SU does not get its first choice SP, it may be happy to receive spectrum from the second choice SP.}.

We use the DA Algorithm to perform the spectrum matching between a set of SUs and SPs.
Given that an SP $m\in \mathcal{M}$ has strict preference for every SU $n\in \mathcal{N}$ and vice versa, each SU makes an offer to its most preferred SP. Each SP accepts the offer from the SU with the highest preference level at its preference list and rejects all other proposals. However, this acceptance is conditionally \emph{held} by the SP until it receives an offer from a more preferred SU. In the next round, all the SUs rejected by the SPs in the previous round would make an offer to the SPs next in line in their preference lists. The SP ${m}\in \mathcal{M}$ will accept the offer from an SU $\hat{n}\in \mathcal{N}$ only if the new proposal from SU $\hat{n}$ is higher at its preference list than the currently held proposal by SU $n$, and reject it otherwise. If the SP accepts the new proposal, it will release the previously \emph{held} offer and inform SU $n$ that it has released its proposal. SU $n$ then proposes to the SP next in its preference list in the next round. This process continues until all the SUs reach to the end of their proposal list. At the end of the process, we have the stable matchings for the scenario discussed.
The pseudocode for the algorithm is shown in Algorithm \ref{algorithm}. For convenience, the frequently used notation is summarized in Table \ref{tab:notation}.

\begin{table}
\begin{center}
\footnotesize
\caption{Notation Summary}
\begin{tabular}{lr}
\toprule
Notation & Definition\\
\midrule
$\mathcal{M}$& Set of Spectrum Providers\\
$\mathcal{N}$&Set of Spectrum Users\\
$P(n)$& Ordered list of preference for SU $n$ over $\mathcal{M}$\\
$P(m)$& Ordered list of preference for SP $m$ over $\mathcal{N}$ \\
$\mu(.)$& One-to-One correspondence for 2 elements in $\mathcal{M}$ and $\mathcal{N}$\\
$S_{n,i}$& $i^{\rm {th}}$ matching success for an SU $n$\\
\bottomrule
\end{tabular}
\vspace{-0.4cm}
\label{tab:notation}
\end{center}
\end{table}

\textbf{Convergence and Complexity: }Convergence for the DA algorithm is guaranteed. As every proposing SU can at best propose to an SP once regardless of the decision (accept/reject) of the SP, the algorithm's convergence is guaranteed after finite iterations. As there are $N$ proposing SUs and in each iteration $M$ SPs are available to be proposed, the computational complexity of the algorithm is $\mathcal{O}(MN)$ provided that preference lists for all $n\in \mathcal{N}$ and $m\in \mathcal{M}$ are a-priori available.

\textbf{Stability: } The DA algorithm always results into a \emph{stable matching} solution, which implies that after matching, there is no SP $m$ which has a higher preference for an SU $n$ and the SU $n$ has higher preference for SP $m$ and they are not matched \cite{Gao:TMC2016}. The stable matchings in general depend on the fact that which entity is making the proposal and which one is accepting/refusing it. In this work, SUs propose first to the SPs to get spectrum which results in specific stable matchings. It is clear that SUs start proposing in the order of highest preference and if accepted are matched to their most preferred SP. Similarly, SPs hold their most preferred SU until they find a better proposal. Both of these steps guarantee that the solution results in stable matching for our spectrum sharing scenario.

\begin{algorithm}[t]
\caption{DA Algorithm for Spectrum Matching}
\KwIn{$(\mathcal{M},\mathcal{N},\textbf{P}$)\;}
$\textbf{P}$ is a matrix, with one row of $P(m)$ for every SP $m$\;
\tcc{T spectrum allocation instants.}
\For{t=1 \KwTo T}{
Generate random $P(n)$ for each SU $n$\;
Initialize all $m \in \mathcal{M}$ and $n \in \mathcal{N}$ to unmatched pairs.\
\tcc{$\mathbf{Id_n}$ is a vector containing id numbers of the SPs which rejected SU $n$'s request}\
$\mathbf{Id_n}=\phi,\forall n$\;
\While {No offer is made by any SU}{
Each SU requests to its preferred SP who
has not rejected its offer yet\;
Each SP holds onto its most-preferred and acceptable (if any) offer and rejects all others\;
\tcc{Each SU records the SP ids who rejected its requests.}\
Update $\mathbf{Id_n}$, $\forall n$\;
}
Match each SP to the SU whose offer it is holding\;
Record the matchings of the SU-SP pairs for instant $t$.
}
Average the matching statistics $S_{i,n},\forall n$ over all $t$.
\KwOut{$S_{i,n},\forall n$;}
\label{algorithm}
\end{algorithm}

\subsection{Many-to-One Scenario}
\begin{figure}[t]
\centering
\includegraphics[width=1.5in]{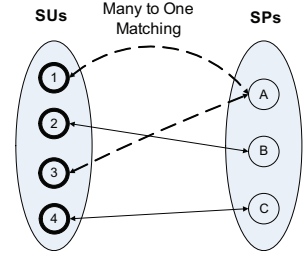}			
\caption {Many-to-one scenario for $M=3$ and $N=4$ where SP A has been shown to be connected with two different SUs.}
\label{fig:MTO}
\end{figure}

In the many-to-one scenario, some or all of the SPs are allowed to have more than one slice of spectrum available which they can allocate to different SUs. Note that every SU is still allowed to have only one slice allocated. Depending on resources available, every SU still can get some spectrum even if $N>M$. This scenario has been depicted in Fig.~\ref{fig:MTO}.

To define the many-to-one matching for the two sided matching market, we first define the notion of quota $q_m$ for an element $m\in \mathcal{M}$. $q_m$ is the cardinality of the set $\mu(m)$ which contains the elements that can be matched with $m\in {\mathcal{M}}$, i.e., $\{\mu(m)\}_{1\times q_m}$. For this work, $q_m$ denotes the maximum number of spectrum slices available at a single spectrum allocation instant for SP $m$. Let us define many-to-one matching in the following.
\begin{definition}{\textbf{Many-to-One Matching:}}
A Many-to-one matching between two disjoint sets $\mathcal{M}$ and $\mathcal{N}$ can be represented by a many-to-one correspondence such that,
\begin{itemize}
  \item $\mu(n)=\{m\}$ if SU $n$ is matched to a spectrum slice from SP $m$,
  \item $\mu(m)\subset \mathcal{N}$ and $|\mu(m)|\leq q_m,\forall m$ and
  \item $\mu(n)=\{m\}$ iff $n\in \mu(m),\forall n,m$.
\end{itemize}
i.e., $m \in \mathcal{M}$ is mapped to $n\in \mathcal{N}$ if and only if $n$ is matched to $m$; and $n$ is one of the $q_m$ elements matched with $m$.
\end{definition}
Thus, many-to-one matching is an extension of one-to-one matching such that each SU $n$ is connected to only one SP (as in one-to-one matching), but an SP $m$ can be connected to multiple SUs with maximum connections bounded by $q_m$.

To solve the problem for many-to-one scenario, we apply the Gale-Shapley algorithm which was used before for the college admission problem. This algorithm is a modification of the DA algorithm presented in Algorithm \ref{algorithm} in such a way that every SP can hold (and finally accept) up to $q_m$ SUs instead of just one as was the case for one-to-one matching. The algorithm terminates when either all $q_m, \forall m$ spectrum chunks have been allocated or no unallocated SU is left. This algorithm also guarantees convergence and stability as the DA algorithm for one-to-one case.

\section{Spectrum Sharing Scenario}
\label{sect:spectrum sharing}
\begin{figure}[t]
\centering
\includegraphics[width=3.5in,trim={0 4cm 0 6cm},clip]{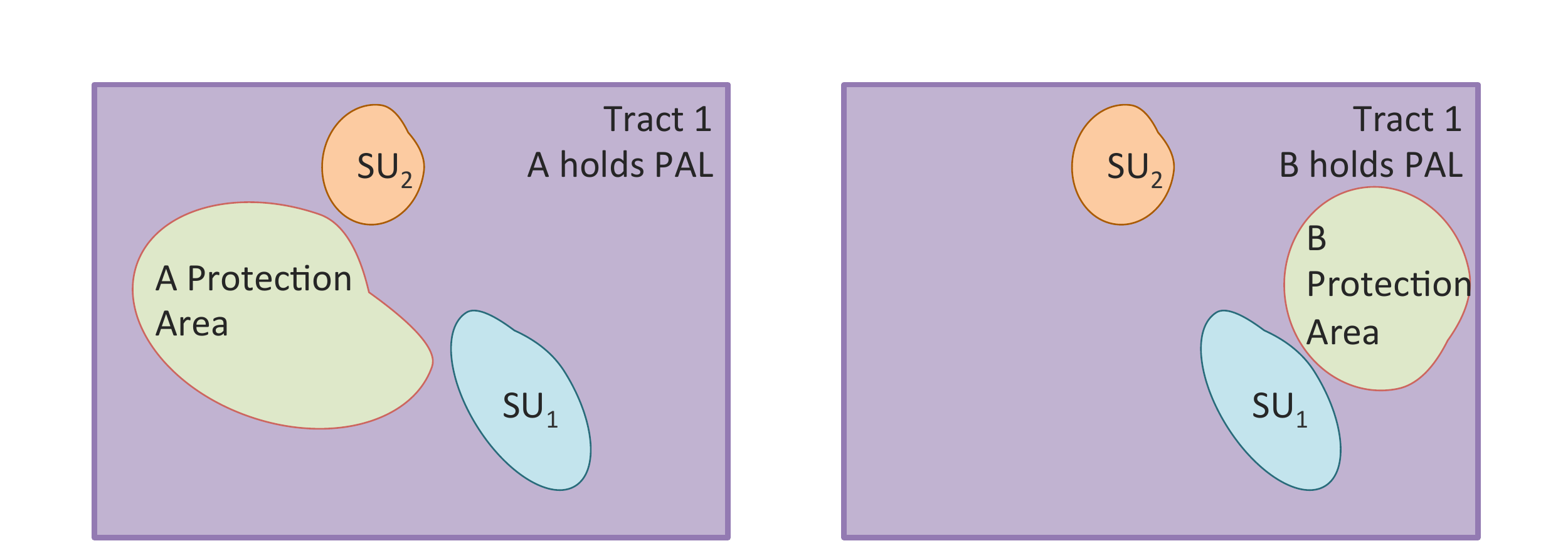}	
\caption {PALs holders will be allowed to lease any bandwidth for any period of time and for any portion of their licensed geographic area within the scope of the PAL but outside of the PAL Protection Area. Green areas represent the Protection Areas of two PAL holders in the same census tract. The purple area is the region in which spectrum can be leased to SUs. The orange and blue areas represent the area of interests of two SUs. For example, SU2 might prefer to lease spectrum from PA Licensee B because of the larger distance between their area of operation. For the same reason, for SU1 the two PA Licensees would be equivalent.}
\label{fig:PALs}

\end{figure}

We evaluate the proposed approach in the context of the CBRS. CBRS is based on a three-tiered sharing framework in which incumbent users - federal and non-federal - represent the highest tier and are protected from interference generated by the two lower tiers - Priority Access (PA) and General Authorized Access (GAA) \cite{Elma:GC2016}. A Priority Access License (PAL) is defined as the authorization to use a $10$ MHz channel in a single census tract for
three years. In particular, PA users will be protected from interference generated by GAA use, while GAA users will receive no interference protection. PA users will be protected along the contour of the PAL Protection Area. Around each deployed Citizens Broadband Service Device (CBSD) a default  protection contour will be determined based on a signal strength of $-96$ dBm in $10$ MHz. PA Licensees may opt to reduce their Protection Area. In fact, PA Licensees may enter into spectrum leasing arrangements with approved entities for areas that are within their Service Area -the census tracts where they have a PAL- and outside of their Protection Areas. Figure \ref{fig:PALs} shows a census tract and two PA Protection Areas (in green) corresponding to two PA Licensees. SUs that are interested in acquiring spectrum resources exclusively in a certain geographic area may interact with one or more PA Licensees to negotiate a leasing arrangement. Since different PA users have potentially different protection areas in the same census tract, each SU can rank PA Licensees depending for example on the size of the available area, on the match between the area of interest (orange and blue areas in Figure \ref{fig:PALs}) and the available area, or on the distance between the area of interest and the PA protection area. Although leasing agreements can be negotiated individually, in this paper we assume that they will be arranged in a common secondary market in which different PA Licensees and SUs can express their preferences.

On the other side, PA Licensees could rank SUs based on different criteria. For example,
\begin{enumerate}
    \item  An MBB
operator holding a PAL might prefer to lease spectrum to a smart grid (SG) or a vehicle to infrastructure (V2X) operator rather than to a direct cellular competitor (target market criterion).
  \item Another criterion might be based on the capability of the SUs to reduce interference. This can be achieved in several ways including spectrum shaping, interference cancellation and beamforming techniques, among others \cite{Hong:2009,Paisana:2015,macaluso2014spectrum}. For example, PA Licensees could rank SUs depending on whether they can avail of transmit beamforming capabilities or not. SUs that are able to perform beamforming would be in general preferred because they would enable the PA Licensee to enter into multiple leasing arrangements in the same Service Area, as a consequence of the inter-SU interference mitigation enabled by beamforming.
\end{enumerate}

To illustrate the potential of matching theory applied to CBRS secondary market, we provide an example use case.
We consider a system with 3 PA Licensees, denoted by A, B and C, which provide services for MBB, SG and V2X applications, respectively. On the other side, we have 4 SUs, denoted by SU1, SU2, SU3 and SU4, such that SU1 and SU2 provide SG services, while SU3 and SU4 provide MBB and V2X services, respectively. SU1 and SU3 have beamforming capabilities while SU2 and SU4 have no such capabilities. Target service areas and hardware capabilities of the networks have been summarized in Table \ref{table_scenario}.
\begin{table}[!t]
\renewcommand{\arraystretch}{1}
\caption{Summary of Service Areas and Hardware Capabilities}
\label{table_scenario}
\begin{center}
\begin{tabular}{|c|c||c|c|c|}
\hline
\multicolumn{2}{|c||}{SP}&\multicolumn{3}{|c|}{SU}\\
\hline
\hline
Identifier & Service Area &  Identifier& Service Area & Beamforming\\
 & & &  &Capability\\
\hline
\hline
A &MBB&1&SG&Yes\\
B&SG&2&SG&No\\
C&V2X&3&MBB&Yes\\
-&-&4&V2X&No\\
 \hline
\end{tabular}
\end{center}
\end{table}

Assuming that the target market criterion takes precedence, priorities of PA licensees would be,
\begin{eqnarray}
\label{eqn:NOpreferences}
P(A)&=&(n_1,n_2,n_4,n_3)\\
P(B)&=&(n_3,n_4,n_1,n_2)\nonumber\\
P(C)&=&(n_1,n_3,n_2,n_4)\nonumber
\end{eqnarray}
PA Licensee A has the least priority for SU3, which provides broadband services and is a competitor to PA Licensee A. The priority for other SUs is based on the interference mitigation criterion for PA Licensee A. As SU2 and SU4 have no beamforming capabilities, they are prioritized just ahead of SU3 in arbitrary order. SU1 is not a competitor and has beamforming capabilities and preferred as first choice by PA Licensee A. The preferences for other PA Licensees follow the same logic.

As an example, based on protection area and their preferred service area, the preferences of SUs can, for example, be defined by,
\begin{eqnarray}
P(1)=(m_A,m_B,m_C)\\
P(2)=(m_B,m_A,m_C)\nonumber\\
P(3)=(m_B,m_C,m_A)\nonumber\\
P(4)=(m_A,m_C,m_B)\nonumber
\end{eqnarray}
where SU1 prioritizes PA Licensee A over other PA Licensees as it provides SU1 a larger service area.

We apply one-to-one matching theory algorithm on this example. SU1 and SU4 prefer PA Licensee A. PA Licensee A prefers SU1 over SU4 and therefore, holds on to the proposal of SU1. SU2 and SU3 both prefer PA Licensee B, but PA Licensee B prefers SU3 over SU2 and holds onto proposal of SU3. SU2 proposes PA Licensee A then, but it prefers already held SU1 over it and rejects the proposal. SU4 proposes to PA Licensee C who holds on to its proposal. Then, SU2 proposes to PA Licensee C who accepts the proposal by releasing SU4 because it has higher priority for SU2 as compared to SU4. Finally, SU4 proposes PA Licensee B who rejects its proposal because it already holds more preferred proposal from SU3. At the termination of the algorithm, all the PA Licensees are matched with the SUs they hold while SU4 remains unallocated.

\section{Performance Evaluation}
\label{sect:performance evaluation}
To investigate the behavior of the proposed matching theory approach, we conduct extensive simulations in which the preferences of the SUs are randomly changed to simulate various combinations of SU preferences resulting into different stable matchings. Since SU preferences depend on distance from the PA Licensee protection area, they can change over time. In fact, while the PA Licensee protection areas stay the same, the SUs' areas of interest can vary due to traffic conditions. For example, SUs could be generally operating as GAA, and temporarily be interested in acquiring spectrum resources and protection from interference in certain geographic areas in case of special events. We assume that the preferences of the SPs are fixed throughout the simulation period because the SUs' beamforming capabilities and the target market remain unchanged.

\subsection{One-to-One Matching}
For the one-to-one scenario, we fix $M=3$ and $N=3$. At a spectrum allocation instant, the SUs have a preference list for their preferred SPs to get a spectrum slice. To focus more on the effect of SP preference list, the SU preferences are randomized with uniform probability distribution to simulate all possible combinations of preferences for the matching scenarios. For example, SU1 may have a preference list $(m_B,m_C,m_A)$ where the order of preference has been chosen from a uniform probability distribution. In the next spectrum allocation instant, another random preference list is generated from the uniform probability distribution.

\begin{figure}[!t]
\includegraphics[width=3.5in]{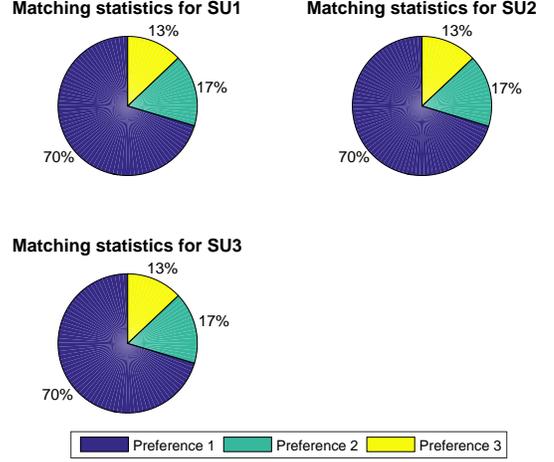}	
\caption {One-to-one matching statistics for $M=3$ and $N=3$.}
\label{fig:OTO33}
\end{figure}

\begin{table}[!t]
\caption{Preferences of SPs for $N=3,M=3$}
    \label{table:preference1to1}
\begin{center}
 \begin{tabular}{ |l|l| }
  \hline
SP&Preference List \\ \hline\hline
$P(A)$ & $(n_1,n_2,n_3)$   \\ \hline
$P(B)$&$(n_2,n_3,n_1)$ \\
\hline
$P(C)$&$(n_3,n_1,n_2)$\\
\hline
 \end{tabular}
\end{center}
\end{table}		

For simulations, the preferences of SPs for SUs are (arbitrary but) fixed, and are summarized in Table \ref{table:preference1to1}. The preferences of SPs have been chosen in such a way that there is a symmetry in overall preference statistics such that every SU appears to be first, second and third preference of exactly one SP.\footnote{Please note that choice of preference list in Table \ref{table:preference1to1} is arbitrary for numerical simulations. In practical systems, the (SP as well as SU) preference lists are generated from the specific algorithms that represent utility gain of the respective SP/SU for each candidate SU/SP.} For example, SU1 is first preference of SP A, third preference of SP B and second preference of SP C.
We perform Monte Carlo simulations to get matching statistics $S_{i,n},\forall i,n$. $10^5$ spectrum allocation instants are simulated with a random preference order generated at each instant for each SU. We average $S_{i,n},\forall i,n$ over $10^5$ simulations to get our results.

Fig. \ref{fig:OTO33} shows pie charts for $S_{n,i}$ for the 3 SUs. As the preferences for the SUs are randomized and the preferences from the SPs are symmetric, the statistics for all the $S_{n,i}, \forall n$ are symmetric. All the SUs get spectrum from their first, second and third choice SP approximately 70\%, 17\% and 13\% times, respectively.

\begin{figure}[!t]
\includegraphics[width=3.5in]{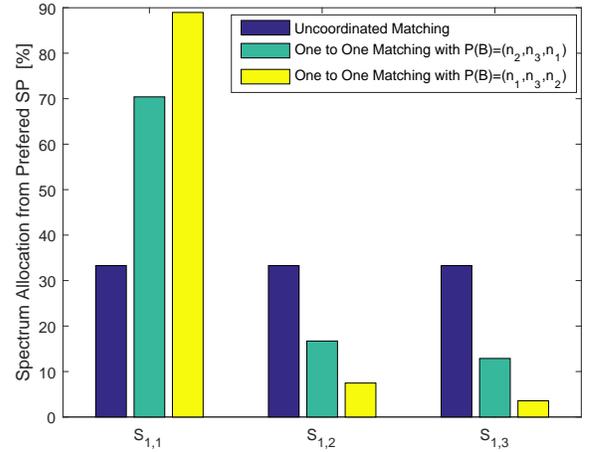}	
\caption {Comparison of one-to-one matching with difference SP preferences and uncoordinated matching.}
\label{fig:OTOcomp}
\end{figure}

\begin{figure*}[!t]
 \centering
 \subfigure[SP Preferences in Table \ref{table:preferencemanyto1}]
 {\includegraphics[width=3.5in]{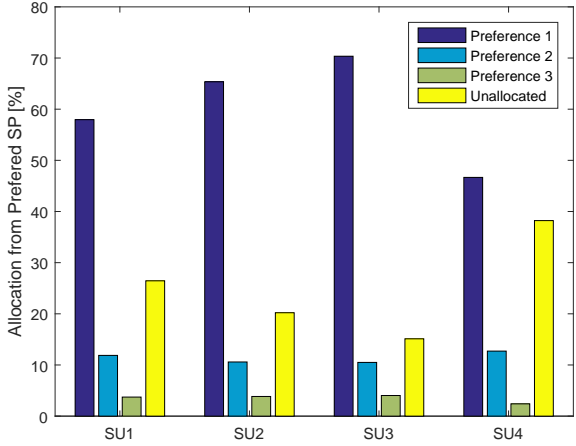}
 \label{fig:pref1}}
\subfigure[SP Preferences in Eq. (\ref{eqn:NOpreferences})]
 {\includegraphics[width=3.5in]{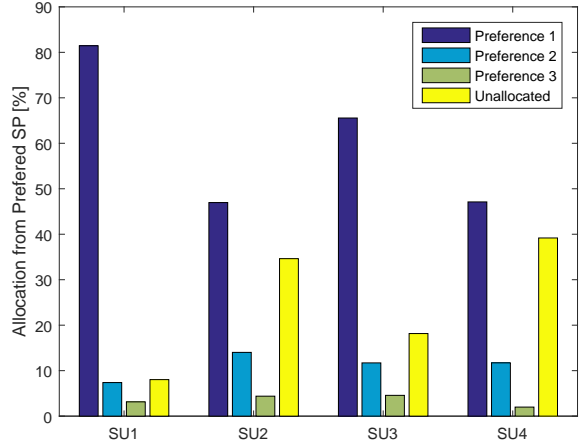}
 \label{fig:pref2}}
   \caption{Comparison of matching statistics for the DA algorithm for 2 different SP preferences.}
  \label{fig:comparison}
\end{figure*}

Fig. \ref{fig:OTOcomp} compares spectrum allocation statistics for $n=1$ for matching theory and uncoordinated (randomized) matching.
As the $S_{n,i}$ statistics for all SUs are the same, we plot statistics for SU1 only. We compare $S_{1,i}$ for SU1 for the DA one-to-one matching algorithm and uncoordinated matching. $S_{1,1}$ is $70\%$ for the DA algorithm, which is only $33\%$ for uncoordinated matching. To study the effect of SP preferences, we change the preference of SP B from $(n_2,n_3,n_1)$ to $(n_1,n_3,n_2)$, thereby giving more preference to SU1. $S_{1,1}$ improves considerably for SU1 from $70\%$ to $89\%$. This explains the underlying business model that if SU1 wants to get spectrum from the SP of its choice, it has to invest in improving its capabilities for interference mitigation.

\begin{table}[!t]
\caption{Preferences of SPs for $N=4,M=3$}
    \label{table:preferencemanyto1}
\begin{center}
 \begin{tabular}{ |l|l| }
  \hline
SP&Preference List \\ \hline\hline
$P(A)$ & $(n_1,n_2,n_3,n_4)$   \\ \hline
$P(B)$&$(n_2,n_3,n_4,n_1)$ \\
\hline
$P(C)$&$(n_3,n_4,n_1,n_2)$\\
\hline
 \end{tabular}
\end{center}
\end{table}

Fig. \ref{fig:comparison} shows statistics for One-to-One matching for $N=3$ and $M=4$ for two different SP preferences. The preferences of the SUs are random as before. The results in Fig. \ref{fig:pref1} and Fig. \ref{fig:pref2} have been obtained using 2 different sets of SP preferences; one summarized in Table \ref{table:preferencemanyto1} and the other in (\ref{eqn:NOpreferences}), respectively. Due to one-to-one matching, we can see that the SUs do not get any spectrum for a considerable percentage of time. As $N$ increases for the SUs, there will be more instants when spectrum slice will not be allocated to the SUs if only one-to-one matching is allowed. SU1 has higher unallocated spectrum instants in Fig. \ref{fig:pref1} as compared to the instances in Fig. \ref{fig:pref2} because it is first choice of one SP in the former case and two SPs in the latter. The opposite effect can be observed for SU2, while SU3 and SU4 behave almost identically due to smaller changes in their relative preference levels in the SP priority lists.

\subsection{Many-to-One Matching}
\begin{figure}[!t]
\includegraphics[width=3.5in]{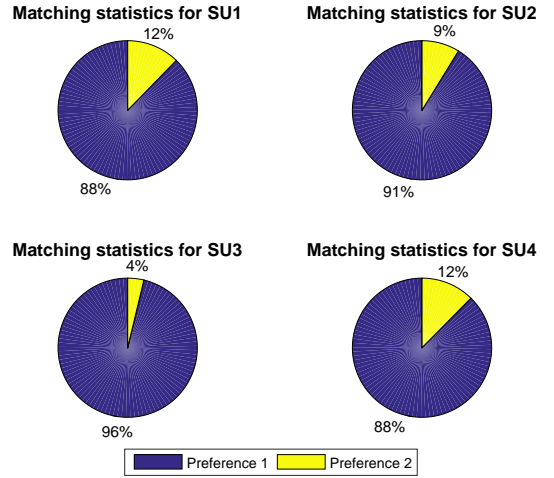}	
\caption {Many-to-one matching statistics for $M=3$ and $N=4$. }
\label{fig:MTO34}
\end{figure}

We evaluate performance of the many-to-one matching algorithm and show how it improves the matching statistics as compared to one-to-one matching, and minimizes probability of unallocated spectrum for SU when $M<N$.
Fig.~\ref{fig:MTO34} shows matching statistics for $M=3$ and $N=4$ case when every SP $m$ can provide one spectrum slice each to at most 2 different SUs at every spectrum allocation instant, i.e., quota $q_m$ equals 2 for each SP $m$. We use SP preference list from Table \ref{table:preferencemanyto1}. It is apparent this results in great improvement in spectrum allocation for all the SUs as compared to one-to-one matching and unallocated spectrum slots vanish completely while the SUs have more opportunities to get spectrum slice of their first and second choice at the same time.

\begin{figure}[!t]
\includegraphics[width=3.5in]{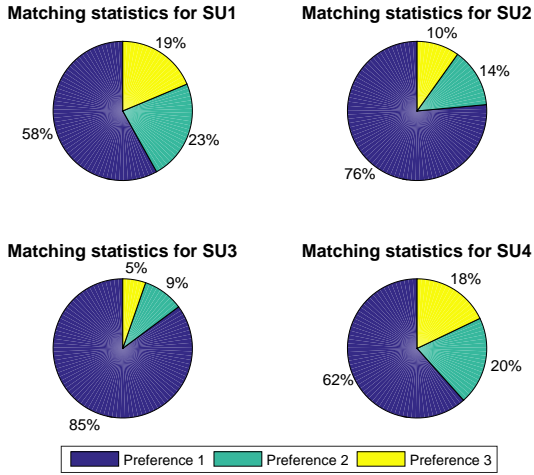}
\caption {Many-to-one matching statistics with only SP A has two slices available.}
\label{fig:MTO34A}
\end{figure}

Fig. \ref{fig:MTO34A} shows matching statistics for the Many-to-One matching for $M=3$ and $N=4$ case, when only SP A have 2 slices available for 2 different SUs, while other SPs have only one slice available. SU2 is the real beneficiary of this 'bias' in spectrum availability from different SPs. SP A has two spectrum slices available, which implies that the SUs who are first and second preference of SP A, will always get spectrum of their first choice if they request spectrum from SP A as their first choice. The SUs who are second preference of SP B and C do not get spectrum slice from SP B and C with probability one even if they prefer SP B and C.

\begin{figure}
\includegraphics[width=3.5in]{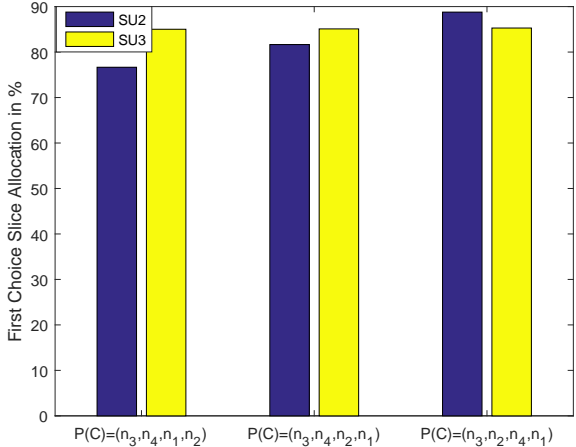}	
\caption {Many-to-one matching statistics with only SP A has two slices available and preference for SP C varies. }
\label{fig:MTOcomp}
\end{figure}

Due to lack of spectrum availability in the market (not every SP can accommodate 2 SUs), the probability of SUs getting spectrum from the SP of their first and second choice decreases as compared to the case discussed in Fig. \ref{fig:MTO34}. This 'bias' in spectrum availability from different SPs shows that the SUs have to adapt their spectrum leasing strategy by investing in a way that improves their chances to lease spectrum from the SP which allows multiple SUs. We observe that SU3 gets slice of its first choice for more time as compared to SU2. To study this behaviour further, we plot the results in Fig. \ref{fig:MTOcomp}.

Fig. \ref{fig:MTOcomp} shows matching statistics for Many-to-One matching for $M=3$ and $N=4$ case, when only SP A has 2 slices available. However, we change preferences of SP C and study the effect on the matching statistics of SU2 and SU3. All other SPs have the same preferences as shown in Table \ref{table:preferencemanyto1}. The reason for SU3 having higher probability to get its first choice (and not SU2) lies in overall distribution of preferences of SU3 for all SPs. SU3 is first preference of SP C, second preference of SP B and third preference of SP A. However, SP A has two spectrum slices available. If one of SU1 and SU2 does not demand a slice from SP A and SU3 does, it gets it. This implies that that SU3 is effectively second preference of SP A. On the contrary, SU2 is second preference of SP A, first preference of SP B; but 4th preference of SP C. If SU2 demands a spectrum slice from SP C as its first preference, it has very small chance of getting it because it requires SUs 1, 3 and 4 not asking for it. This has a huge impact on overall statistics for SU2 to get spectrum of its first preference. To quantify this effect, we repeat the experiment by changing preferences of SP C with respect to SU2. We do not change preferences for SU3. We observe, when SU2 moves ahead in the preference list of SP C, its statistics start improving in terms of getting spectrum slice of its first preference. When SU2 moves to be the second preference of SP C, its $S_{2,1}$ is higher than $S_{3,1}$ in spite of being first preference of only SP B. This shows that it is not necessary for an SU to be first choice of more SPs to increase its chances of getting spectrum of first chance. It should be noted that moving ahead in SP priority list requires to equip with beamforming (costly hardware) capabilities and requires investment from the SU.

\section{Conclusions}
\label{sect:conclusions}
We apply matching theory to a spectrum sharing scenario between the SP and SU networks to benefit both parties. The SPs and SUs have a preference order that depends on their individual benefits from spectrum sharing with each other. We discuss a general framework first and apply it to CBRS as an example of commercial spectrum sharing. We numerically evaluate the performance of matching theory algorithms for one-to-one and many-to-one spectrum sharing scenarios and quantify the effect of preferences of both SPs and SUs. The ability of an SU to get spectrum from the SP is measured by a metric $S_{n,i}$, which is the probability of an SU $n$ to get spectrum from the SP of its $i^{th}$ preference in its preference list. The results show that application of matching theory DA algorithms improves $S_{n,1}$ for the SUs as compared to uncoordinated matching of SP-SU pairs. One of the main advantages of using matching theory for spectrum allocation is the possibility for both SPs and SUs to express preferences that can embed complex considerations not necessarily or exclusively related to technical requirements for network operation. This framework can be very helpful in spectrum sharing scenarios in future networks, where it is critical to match the priorities of various SP and SU networks to make the best use of the spectrum resources.
\section*{Acknowledgements}
We are thankful to Paul Culhane for performing simulation work in part.

\bibliographystyle{IEEEtran}
\bibliography{bibliography_v2}

\end{document}